\documentclass{article}

\usepackage{arxiv}

\usepackage[utf8]{inputenc} 
\usepackage[T1]{fontenc}   
\usepackage{hyperref}
\usepackage{url} 
\usepackage{booktabs}
\usepackage{amsfonts}
\usepackage{nicefrac}
\usepackage{microtype}
\usepackage{graphicx}
\usepackage{xcolor}
\usepackage{comment}
\usepackage{amsmath}

\title{A Robust Spearman Correlation Coefficient Permutation Test}

\author{ 
	Han Yu\thanks{Corresponding author} \\
	Department of Biostatistics and Bioinformatics\\
	Roswell Park Comprehensive Cancer Center\\
	Buffalo, NY 14263 \\
	\texttt{Han.Yu@RoswellPark.org} \\
	\And
	Alan D.~Hutson \\  
	Department of Biostatistics and Bioinformatics\\
	Roswell Park Comprehensive Cancer Center\\
	Buffalo, NY 14263 \\
	\texttt{Alan.Hutson@RoswellPark.org} \\		
}

\date{}

\renewcommand{\headeright}{}
\renewcommand{\undertitle}{}

\hypersetup{
pdftitle={Robust Permutation Test of the Spearman Correlation Coefficient},
pdfsubject={stat.ME, stat.AP}
pdfauthor={Alan D.~Hutson, Han Yu},
pdfkeywords={rank correlation, studentized, small sample, non-normality},
}

\begin{document}
\maketitle

\begin{abstract}
In this work, we show that Spearman's correlation coefficient test about $H_0:\rho_s=0$ found 
in most statistical software packages is  theoretically incorrect and performs poorly when
bivariate normality assumptions are not met or the sample size is small. The historical works about these
tests make an unverifiable assumption that the approximate bivariate normality of original data
justifies using classic approximations. In general, there is common misconception that the tests about $\rho_s=0$ are robust to deviations from bivariate normality. In fact, we found under certain scenarios violation of the bivariate normality assumption has severe effects on type I error control for the most commonly utilized tests. To address this issue, we developed a robust permutation test for testing the general hypothesis $H_0: \rho_s=0$. The proposed test is based on an appropriately studentized statistic. We will show that the test is theoretically asymptotically valid in the general setting when two paired variables are uncorrelated but dependent. This desired property was demonstrated across a range of distributional assumptions and sample sizes in simulation studies, where the proposed test exhibits robust type I error control across a variety of settings, even when the sample size is small. We demonstrated the application of this test in real world examples of transcriptomic data of the TCGA breast cancer patients and a data set of PSA levels and age.
\end{abstract}


\section{Introduction}
The concept of correlation and regression was originally conceived by Galton when studying how strongly the characteristics of one generation of living things manifested in the following generation \cite{stanton2001galton}. The ideas prompting the development of more mathematically rigorous treatment of correlation 
were developed by Karl Pearson in 1896, which yielded the well-known Pearson Product Moment Correlation Coefficient \cite{pearson1896vii} given as
\begin{equation}
\rho(X, Y) = \frac{Cov(X,Y)}{\sigma_X\sigma_Y} = \frac{E(X - \mu_X)(Y - \mu_Y)}{\sqrt{E(X - \mu_X)^2E(Y - \mu_Y)^2}},
\end{equation}
where $X$ and $Y$ are two random variables from a non-degenerative joint distribution $F_{XY}$, $Cov(X,Y)$ denotes the covariance, and $\mu_X$ and $\mu_Y$, $\sigma_X$ and $\sigma_Y$ are the population means and standard deviations, respectively. If we let $(X_1,Y_1)$, $(X_2,Y_2)$, …, $(X_n,Y_n)$ denote $n$ paired i.i.d. observations, then the sample Pearson correlation coefficient is given as
\begin{equation}
\label{eq:sample_pearson}
r(X, Y) = \frac{ \Sigma_{i=1}^n(X_i-\bar{X})(Y_i-\bar{Y}) }{\sqrt{\strut\Sigma_{i=1}^n(X_i-\bar{X})^2}\sqrt{\strut\Sigma_{i=1}^n(Y_i-\bar{Y})^2}}.
\end{equation}

Shortly thereafter Pearson's work was published  Spearman introduced the rank correlation coefficient in 1904, with the advantages of being robust to extreme values and disparities of distributions between two variables \cite{spearman1961proof}.  It should be noted however that
K. Pearson, in his biography of Galton, says that the latter "dealt with the correlation of ranks before he even reached the correlation of variates, i.e. about 1875", but Galton apparently published nothing explicitly \cite{kendall1979advanced}.  Mathematically, Spearman’s correlation coefficient is defined as the Pearson correlation coefficient on the ranks of $(X_1,Y_1)$, $(X_2,Y_2)$, …, $(X_n,Y_n)$ and denoted as
\begin{equation}
\rho_s (X,Y)=\rho(a,b),
\end{equation}
where $a_i= Rank(X_i)$ and $b_i=Rank(Y_i)$ are the ranks of $X_i$ and $Y_i$, respectively, $i=1,2, \cdots, n$.
In general, when discussing the Spearman correlation coefficient little attention is given to its population measure. 
However,  if we consider that   $a_i/n$ converges to $F(X_{i})$, then one may consider the population measure linked to Spearman's sample correlation coefficient as,
\begin{equation}
\rho_s (X,Y)=\frac{E\{F_X(X)-E[F_X(X)]\}\{F_Y(Y)-E[F_Y(Y)] \}}{E\{F_X(X)-E[F_X(X)]\}^2 E\{F_Y(Y)-E[F_Y(Y)]\}^2}
=\frac{E[F_X(X)-\frac{1}{2}][F_Y(Y)-\frac{1}{2}] }{E[F_X(X)-\frac{1}{2}]^2 E[F_Y(Y)-\frac{1}{2}]^2},
\end{equation}
where $F_X$ and $F_Y$ are the marginal cumulative distribution functions (CDFs) for $X$ and $Y$, respectively. 
The sample estimator of $\rho_s$ can be obtained by replacing the original observations with their ranks in Equation \ref{eq:sample_pearson},
\begin{equation}
\label{eqn:spearman}
r_s(X, Y) = r(a, b) = \frac{ \Sigma_{i=1}^n(a_i-\bar{a})(b_i-\bar{b}) }{\sqrt{\Sigma_{i=1}^n(a_i-\bar{a})^2}\sqrt{\Sigma_{i=1}^n(b_i-\bar{b})^2}} = 1-\frac{6\Sigma_{i=1}^n(a_i - b_i)^2}{n(n^2-1)}.
\end{equation}
For samples from a bivariate normal population, there is also a known relation between the Spearman and Pearson correlation coefficients \cite{moran1948rank}, which is
\begin{equation}
E(r_s) = \frac{6}{\pi (n+1)}\{\sin^{-1}\rho + (n-2)\sin^{-1}\frac{1}{2}\rho\}.
\end{equation}
While
Pearson's $\rho$ measures the linear relationship between two random variables it is often described that Spearman's $\rho_s$ measures a monotonic association
between $X$ and $Y$, thus it may be considered more general measure of association, albeit
it does measure the linear association between $F(X)$ and $F(Y)$. Spearman's correlation coefficient is also less sensitive to extreme values because it is rank based. Due to these advantages, it is widely used as a measure of association between two measurements. It is often of interest to test whether two random variables are correlated, i.e. $H_0:\rho_s=0$, for which the common methods include  a $t$-distribution based test, incorrectly based on the approximate
bivariate normality of the ranks, a test based on Fisher's $Z$ transformation, again assuming approximate bivariate normality of the ranks, 
and what we term the naive permutation test. The $t$-distribution based test  is commonly used when the sample size is large, with the $t$-statistic defined as
$$t=r_s \sqrt{\frac{n-2}{1-r_s^2} }.$$
Under bivariate normality assumptions this statistic approximately follows student’s $t$ distribution with $n-2$ degrees of freedom under $H_0$. For the test based on Fisher’s $Z$ transformation, the statistic is defined as

$$ Z = \frac{1}{2}\arctan(r_s) = \frac{1}{2}\ln \frac{1+r_s}{1-r_s}. $$

Under bivariate normality assumptions the transformed $Z$ statistic approximately follows normal distribution $N(0, \frac{1.06}{n-3})$ under $H_0$. For small sample size scenarios, naive permutation tests are also often used, where $X$ and $Y$ are randomly shuffled separately to simulate the sample distribution of $r_s$ under $H_0$, which is an exact test for testing the independence between $X$ and $Y$ under the null exchangeability assumptions, but may be
an invalid test for testing $H_0:\rho_s=0$ given $G(F(X),F(Y))= G(F(X))G(F(Y))$ does not imply $\rho_s=0$, where $G(\cdot,\cdot)$ denotes
the joint CDF of $F(X),F(Y)$.

These tests are so widely used that they are often the default options in common statistical software packages such as R \cite{team2013r} and SAS \cite{sas2015base}. However, there is little discussion that these tests relies on the untenable assumption that the underlying sample distribution of the ranks
 has a  bivariate normal distribution, which is in fact an impossibility. Even among those who noted this assumption, there is a misconception that the above tests are robust to such deviations because Spearman’s $\rho_s$ is rank based. This is exemplified in a discussion by Feller et al. in their article \cite{fieller1957tests},
\begin{quote}
Conversely, starting from any bivariate distribution $\phi(X,Y)$ we can always find monotonic transformations $X=f(x)$, $Y=g(y)$ to standardized normal variates $x$ and $y$. The resulting bivariate distribution $\psi(X,Y)$ will not necessarily be bivariate normal, but we think it likely that in practical stations it would not differ greatly from this form. This is a field in which further investigation would be of considerable interest. 
\end{quote}
However, as we will show in Section \ref{sec:simulations}, all the commonly used tests about $\rho_s=0$ as discussed above, including 
the naive permutation test, are not even asymptotically valid when the non-exchangeabilty assumptions are violated
under $H_0$. In some cases, the type I error can severely drift away from the desired level as the sample size increases! A undesirable feature
that is more notable in the era of ``big-data''.   Another variation of this approach is the
  Fisher-Yates coefficient, which transforms the original $X$ and $Y$ to their corresponding normal quantiles before the testing \cite{fisher1938statistical}. Although the marginal distributions are of the transformed variates take a pseudo normal form, the joint normality of these transformed values is not guaranteed.

In terms of our modified permutation test it is important to note the classic large sample result in Serfling where the “distribution free” large sample normal approximation for the sampling distribution for Pearson's sample correlation coefficient $r$
is derived using the multivariate delta method \cite{kendall1979advanced}. This method guarantees type I error converges to $\alpha$ when $n\to \infty$ given finite fourth moments. A straightforward way to obtain a similar result for Spearman's correlation is given by replacing the i.i.d sample with corresponding ranks. The test is asymptotically valid because the ranks are asymptotically independent, as we will discuss in Section \ref{sec:methods}.
Even though large sample approximations about these estimators are asymptotically valid they tend to suffer inflated type I errors in the small sample setting,
e.g. $n<50$.

To address this issue, we propose a studentized permutation test for  Spearman's correlation $\rho_s$, which extends the work or Diccicio and Romano for Pearson’s correlation coefficient
 $\rho$ \cite{diciccio2017robust}. We will show that the proposed test is asymptotically valid under general assumptions and is exact under
exchangeability assumptions when  $G(F(X),F(Y))= G(F(X))G(F(Y))$, i.e. more simply when  $X$ and $Y$
are independent.  
We show that our newly proposed test 
has robust Type I error  controls type I error control even when the sample distribution 
is dependent (non-exchangeability)  and non-normal. Importantly, the type I error is well controlled when the sample size is small. This will be illustrated by a set of simulation studies. Finally, we will demonstrate the application of this test in real world examples of transcriptomic data of TCGA breast cancer patients, as well as a data set of PSA levels and age.

\section{Methods}
\label{sec:methods}

\subsection{Background information}

In this section, we start by reviewing the robust permutation test for Pearson's correlation coefficient as proposed by Diciccio and Romano\cite{diciccio2017robust}.
Towards this end, we define $\textbf{G}_n$ to be the set of all permutations $\pi$ of $\{1, \dots, n\}$. For testing independence between two random variables $X$ and $Y$, the permutation distribution of any given test statistic $T_n(X^n, Y^n)$ is defined as 
\begin{eqnarray}
\label{perm_test}
\hat{R}_n^{T_n}(t) = \frac{1}{n!}\sum_{\pi \in G_n} I\{T_n(X^n, Y^n_{\pi})\le t\}
\end{eqnarray}
where $Y^n_{\pi}$ represents $\{Y_{\pi(1)}, \dots, Y_{\pi(n)}\}$. In this setting, the permutation $\textbf{G}_n$ is all possible pairwise combinations between $X^n$ and $Y^n$. A level $\alpha$ one-sided permutation test rejects if $T_n(X^n, Y^n_{\pi})$ is larger than the $1-\alpha$ quantile of the permutation distribution. The permutation test is exact when exchangeability assumptions hold, that is, the distribution of $ (X^n, Y^n)$ is invariant under the group of transformations $\textbf{G}_n$. The test using the Pearson correlation coefficient $\hat{\rho}$ is exact when used a metric of dependence for testing the null hypothesis of independence given as 
$$ H_0: P=P_X \times P_Y, $$
where $P_X$ and $P_Y$ are marginal distributions of $X$ and $Y$, respectively. The null hypothesis of independence is not equivalent to the test about zero correlation given as $H_0: \rho=0$ with the exception of limiting assumptions such as the data are distributed as bivariate normal random variables. In other words, in the general setting two random variables can be dependent but uncorrelated. In such cases, DiCiccio and Romano \cite{diciccio2017robust} have shown that, with finite fourth moments, the permutation distribution of $\hat{\rho}$ converges to $N(0,1)$, but its sampling distribution converges to $N(0, \tau^2)$, where
$$\tau^2 = \frac{\mu_{22}}{\mu_{20}\mu_{02}},$$ 
and
$$\mu_{pq} = E[(X_1 - \mu_1)^p(Y_1 - \mu_2)^q].$$
Thus the test will not be level $\alpha$ unless $\tau = 1$. In light of this result, DiCiccio and Romano proposed a studentized correlation test statistic, which has been shown to control Type I error asymptotically at $\alpha$ when two random variables are dependent but uncorrelated \cite{diciccio2017robust}. Specifically, the studentized statistic is defined as $S_n = \sqrt{n}\hat{\rho}_n/\hat{\tau}_n$, where 
$$\hat{\tau}_n^2 = \frac{\hat{\mu}_{22}}{\hat{\mu}_{20}\hat{\mu}_{02}},$$
$$\hat{\mu}_{pq} = \frac{1}{n} \sum^n_{i=1}(X_{i} - \bar{X})^p(Y_{i} - \bar{Y})^q.$$
The permutation distribution and sampling distribution of $S_n$ both converge to the standard normal distribution asymptotically. It should be noted that even though the results presented in DiCiccio and Romano\cite{diciccio2017robust} are based on large sample approximations, the behavior of this test for small to moderate sample sizes is quite good as born out in their simulation results.

\subsection{ Spearman's permutation correlation test}

 Spearman's coefficient $\rho_s(X, Y)$ is the Pearson correlation coefficient of the ranks of $X$ and $Y$, that is $\rho(a, b)$. When there are ties in the data, their ranks are typically taken as an  average. Unlike Pearson's correlation coefficient $\rho$, which measures the linear relationship between two random variables,  Spearman's correlation coefficient $\rho_s$ measures a monotonic association, thus is far less restrictive. Note that Spearman's correlation coefficient is
also the linear measure between $F(X)$ and $F(Y)$.
 It is also less sensitive to non-normality or extreme values. 

Despite the above advantages, it is a misconception that the tests of  $\rho_s=0$ based on the bivariate normality assumptions underlying the 
original data
will be robust to the deviation from this assumption. In fact, when the original data are independent, their ranks will will be dependent. Thus, tests of $\rho_s$ typically suffer similar issue as for $\rho$.  We also emphasize that "normality" refers to the \emph{joint} normality 
as  opposed to marginal normality, because two random variables that are marginally normal can have a joint non-normal distribution. Therefore, the Fisher-Yates coefficient, which back transforms a variables rank through the normal quantile function does not provide what heuristically one may
consider as a simple correction. In Section \ref{sec:simulations}, we will empirically show that violation of the joint normality assumption will have 
severe effect on type I error control. In addition, it is in fact impossible for the joint distribution of the ranks to be bivariate normal.

Our approach is to replace $(X_i, Y_i)$ with their ranks $(a_i, b_i)$ in order to develop a Spearman's correlation permutation test analog to the
Pearson's correlation permutation test, with some subtle differences. The studentized permutation test of the Pearson's $\rho$ only requires finite fourth moments and that observations are i.i.d. Although the pairs of ranks $(a_i, b_i)$ are no longer i.i.d observations, we can show that they asymptotically satisfy this condition. 

When $(X_i, Y_i)$ are from paired i.i.d. observations, we have $(F_X(X_i), F_Y(Y_i))$ being i.i.d. as well. Since $a_i=Rank(X_i)=n\hat{F}_X(X_i)$ and $\hat{F}_X(X_i) \to F_X(X_i)$ as $n\to \infty$, we have $\frac{a_i}{n} \to F_X(X_i)$. Therefore, the paired observations $(a_i, b_i)$ are asymptotically i.i.d. Consequently, the exchangeability condition will hold at least asymptotically and the test will be asymptotically exact. Intuitively, when $n$ is sufficiently large, knowing $(a_i, b_i)$ will lend little knowledge on the ranks of another pair $(a_j, b_j)$.  It can also be shown that the correlation between $a_i$ and $a_j (i\ne j)$ is approximately $-\frac{1}{n-1}$. Therefore, for a sample sequence $X^n$, the correlation matrix of the ranks converges to $I_n$ when $n \to \infty$.

Specifically, the one sided studentized permutation test for testing $H_0: \rho_s = 0$ versus $H_1: \rho_s >0$ is performed by the following steps. The test is implemented in the R \verb|perk| (\textbf{per}mutation tests of correlation c\textbf{(k)}oefficients) package, which will be available on CRAN (The Comprehensive R Archive Network, https://cran.r-project.org/) and GitHub (https://github.com/hyu-ub/perk).

\begin{itemize}
	\item For $n$ paired i.i.d. observations $(X_1,Y_1)$, $(X_2,Y_2)$, …, $(X_n,Y_n)$, calculate their ranks within each random variable, $(a_1,b_1)$, $(a_2,b_2)$, …, $(a_n,b_n)$.
	\item Estimate the Spearman's $\rho_s$ using Equation \ref{eqn:spearman} as $r_s$.
	\item Estimate the variance of sample estimates $r_s$ by
	$$\hat{\tau}_n^2 = \frac{\hat{\mu}_{22}}{\hat{\mu}_{20}\hat{\mu}_{02}},$$
	$$\hat{\mu}_{pq} = \frac{1}{n} \sum^n_{i=1}(a_{i} - \bar{a})^p(b_{i} - \bar{b})^q.$$
	\item Calculate the studentized statistic $R_s = r_s/\tau_n$.
	\item Randomly shuffle $(b_1, b_2, ... b_n)$ for $B$ times. For each permutation, calculate the permuted studentized statistic $R_s^k$, $k \in (1, ..., B)$.
	\item Calculate the p-value by $$p=\frac{1}{B}\Sigma^B_{k=1}I(R_s^k>R_s).$$
	\item Reject $H_0$ if $p<\alpha$.
\end{itemize}

\section{Simulations}
\label{sec:simulations}

We examined the Type I error control  across all of the tests introduced above
using distributions commonly found in the literature for these examinations across a wide range of settings \cite{diciccio2017robust, hutson2019robust}. For our simulation study, we focused on testing $H_0: \rho_c = 0$ versus $H_1: \rho_c > 0$, with sample sizes $n = 10, 25, 50, 100, 200$. Each simulation utilized $10,000$ Monte Carlo replications and the number of permutations used is $1,000$. We compared the $t$ test, Fisher's $Z$-transformation (Fisher's $Z$), Fisher-Yates method, Serfling's large sample normal approximation (Asymp Norm), naive permutation test (Permute), and studentized permutation test (Stu Permute). The Type I error control for $\alpha=0.05$ was examined. The simulation scenarios 1 through 5 from DiCiccio and Romano. Two additional distributions were studied as well:

\begin{itemize}
	\item[1.] Multivariate normal (MVN) with mean zero and identity covariance.
	\item[2.] Exponential given as $ (X, Y) = rS^Tu $ where $S = diag(\sqrt{2}, 1)$, $r\sim \exp(1)$, and $u$ is uniformly distributed on the two dimensional unit circle.
	\item[3.] Circular given as the uniform distribution on a two dimensional unit circle.
	\item[4.] $t_{4.1}$ where $X=W+Z$ and $Y=W-Z$, where $W$ and $Z$ are iid $t_{4.1}$ random variables.
	\item[5.] Multivariate $t$-distribution (MVT) with 5 degrees of freedom.
	\item[6.] Mixture of two bivariate normal distributions given as $(X, Y) = WZ_1 + (1-W)Z_2$ where $W \sim Bernoulli(0.5)$, $Z_1 \sim N( \begin{pmatrix} 0 \\0 \end{pmatrix}, \begin{pmatrix} 1 & \rho \\ \rho & 1 \end{pmatrix}) $, $Z_2 \sim N( \begin{pmatrix} 0 \\0 \end{pmatrix}, \begin{pmatrix} 1 & -\rho \\ -\rho & 1 \end{pmatrix}) $. We select a range of $\rho$'s: 0.1, 0.3, 0.6 and 0.9 to simulate different degrees of dependencies between $X$ and $Y$ (MVN 1, MVN 3, MVN6, MVN 9).
	\item[7.] Mixture of four bivariate normal distributions (MVN 45), given as $(X, Y) = \Sigma_{k=1}^4 I(W_i=k)Z_i$ where $P(W=k) = 0.25$, $k=1,2,3,4$. In addition, $Z_k \sim N(\mu_k, I_2) $, where $\mu_1 = (5, 5)^T$, $\mu_2 = (5, -5)^T$, $\mu_3 = (-5, 5)^T$, $\mu_4 = (-5, -5)^T$.
\end{itemize}

\begin{figure}[h]
	\centering
	\includegraphics[width=0.2\textwidth]{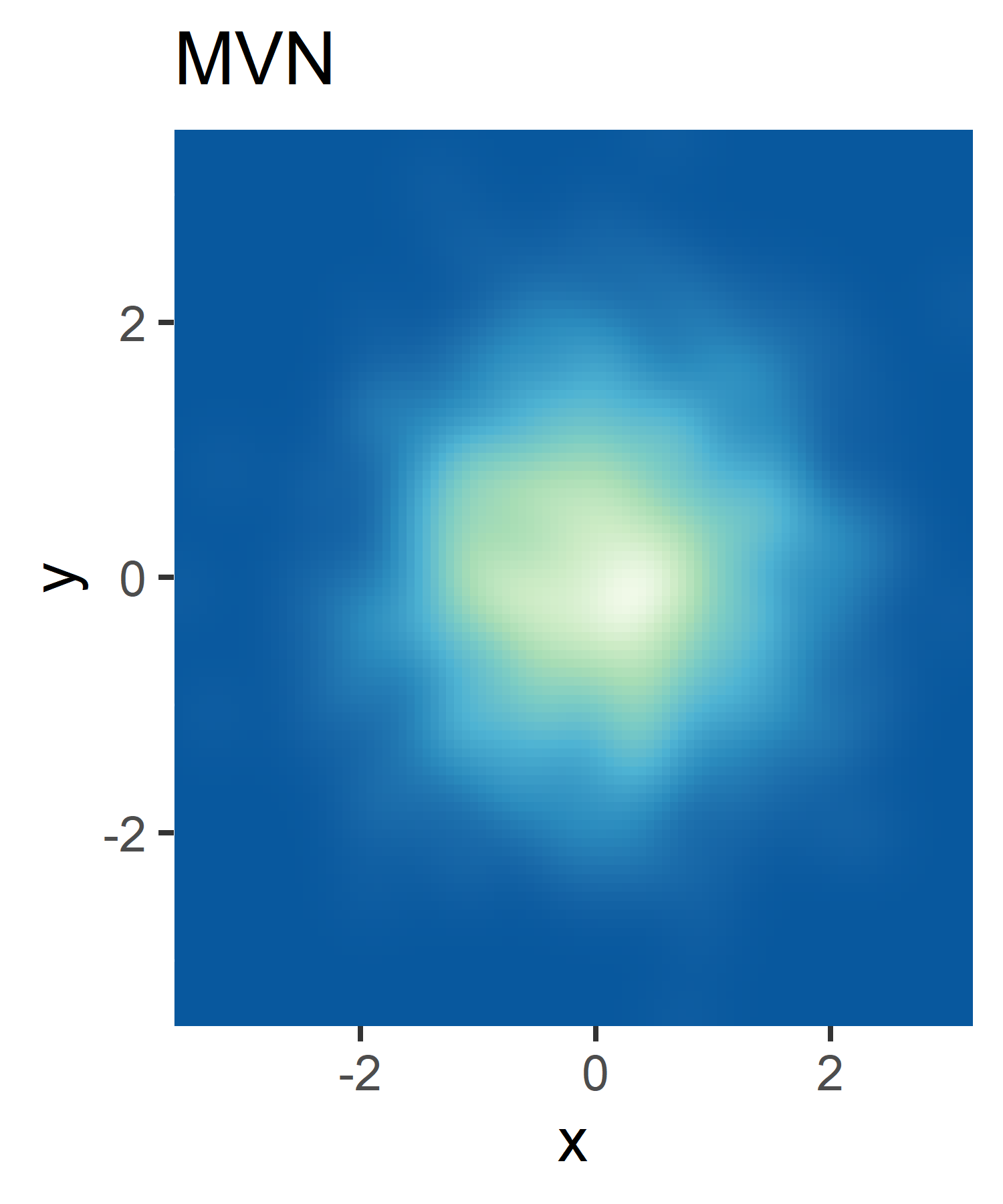}
	\includegraphics[width=0.2\textwidth]{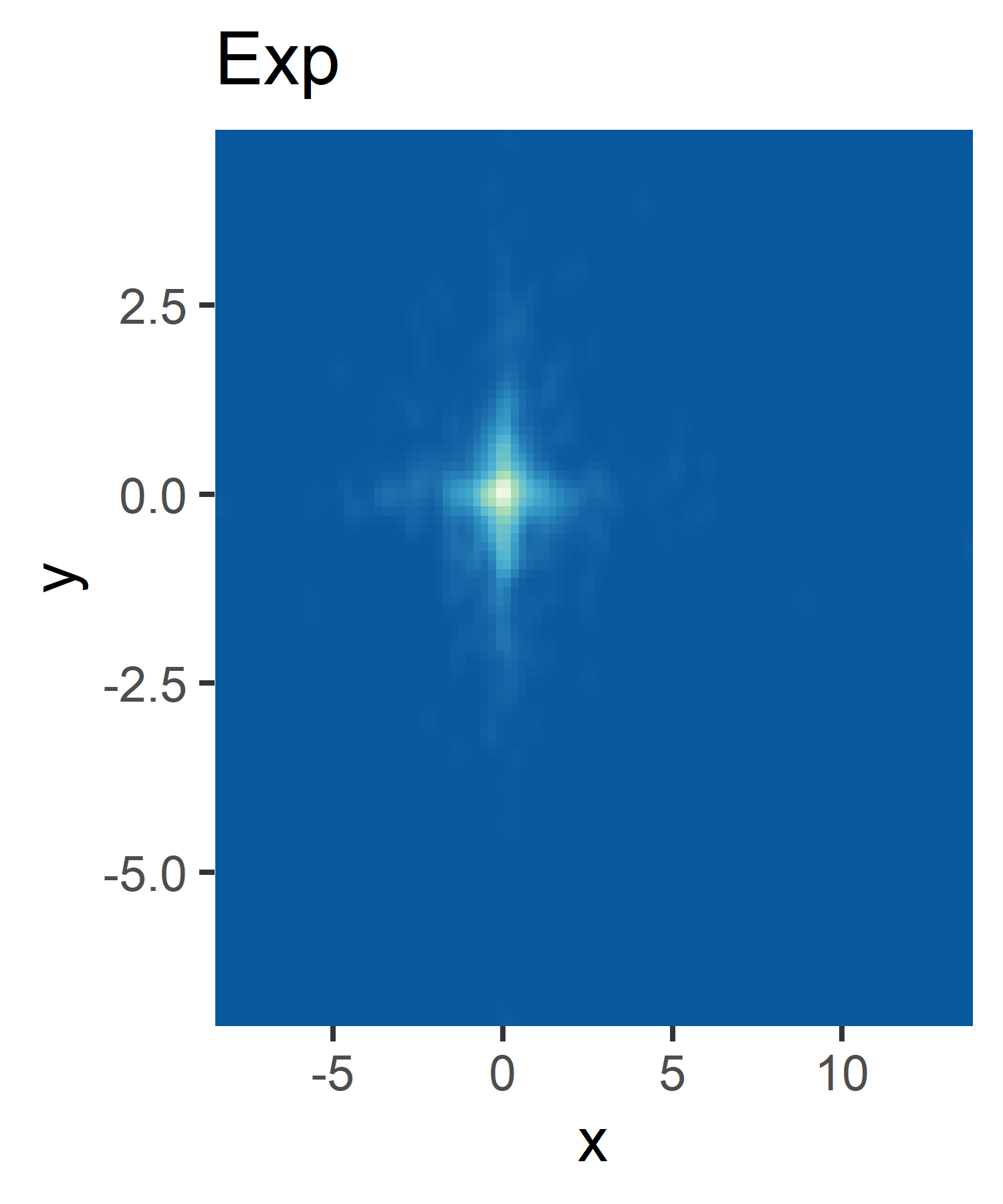}
	\includegraphics[width=0.2\textwidth]{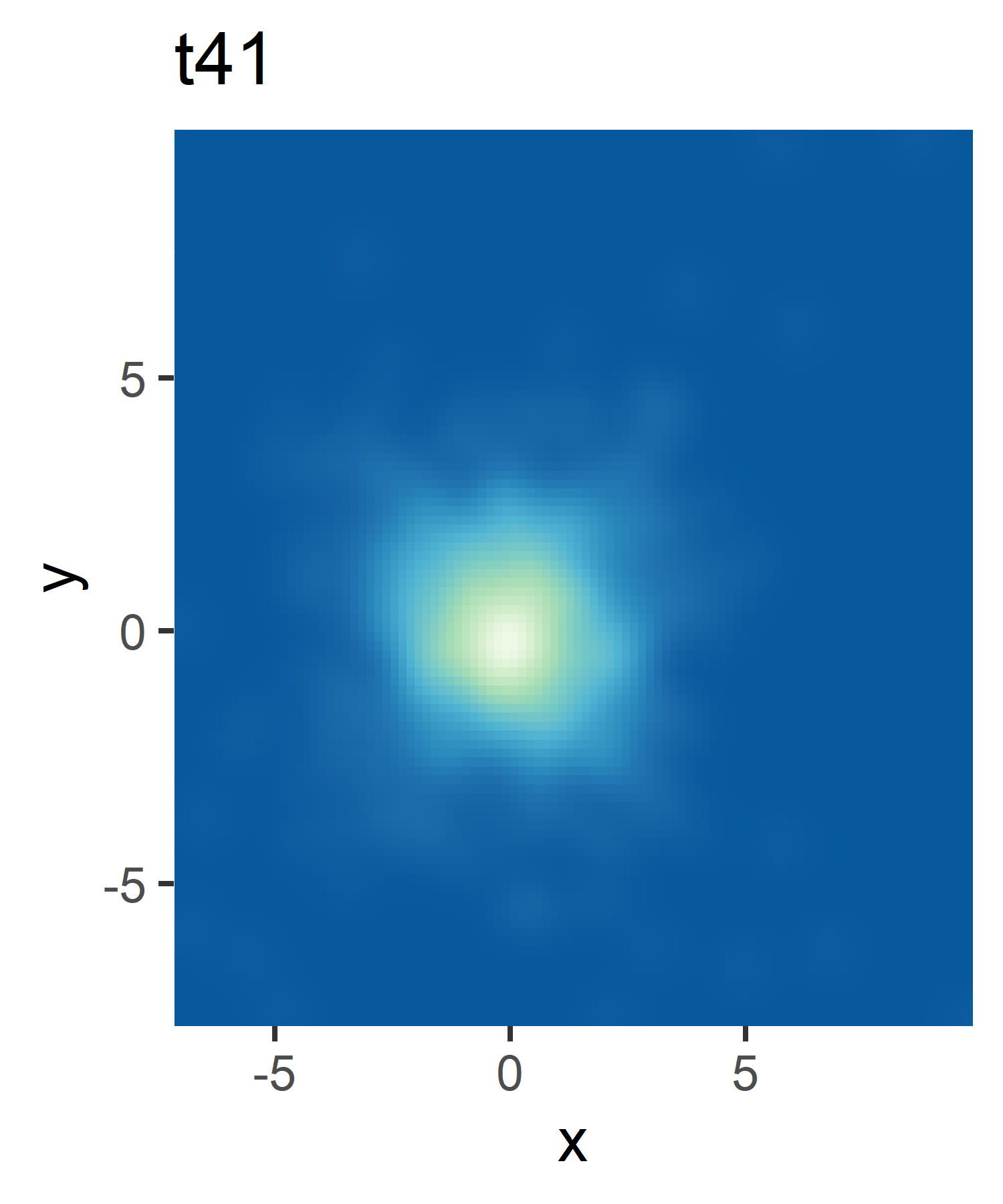}
	\includegraphics[width=0.2\textwidth]{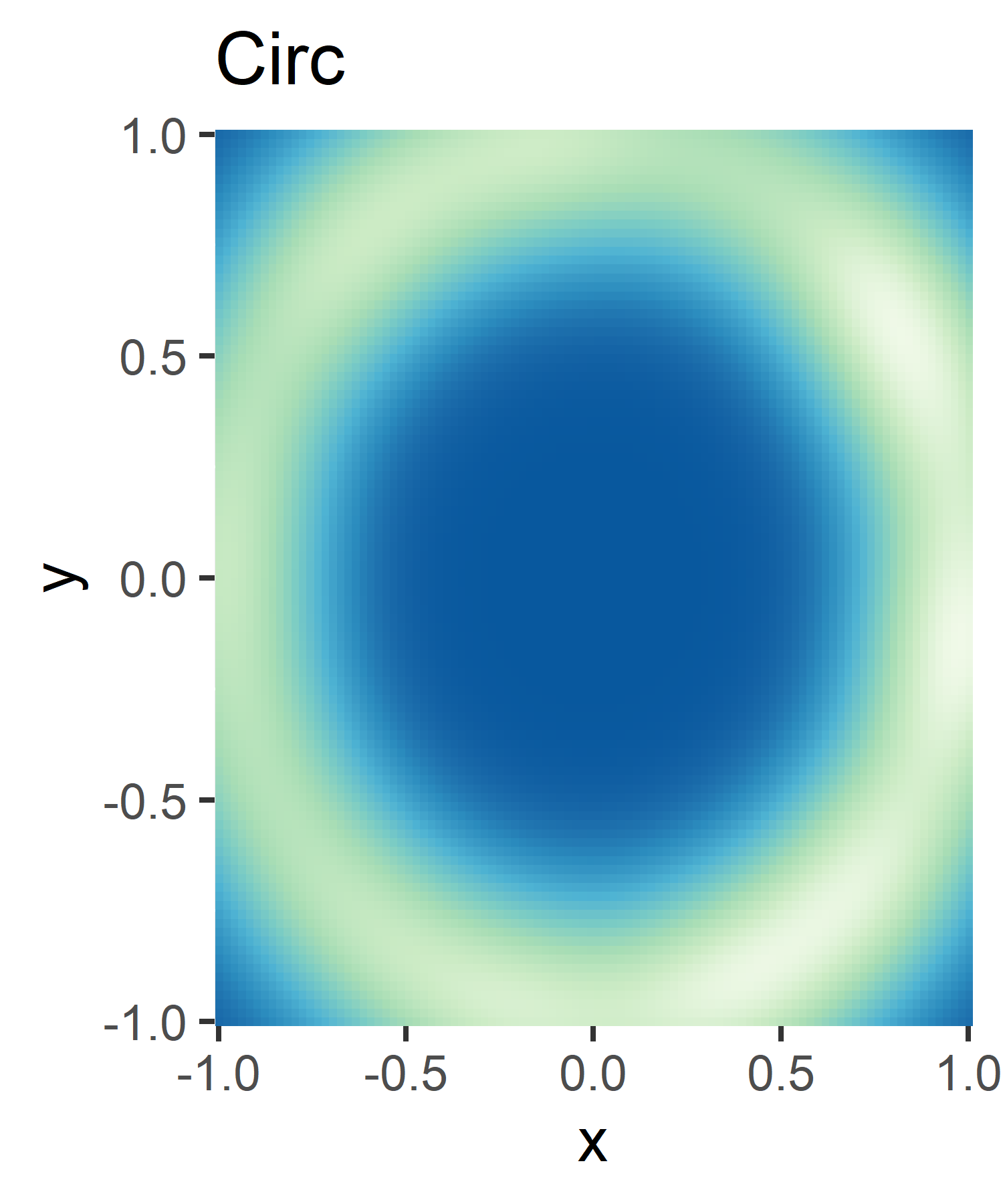}
	\includegraphics[width=0.2\textwidth]{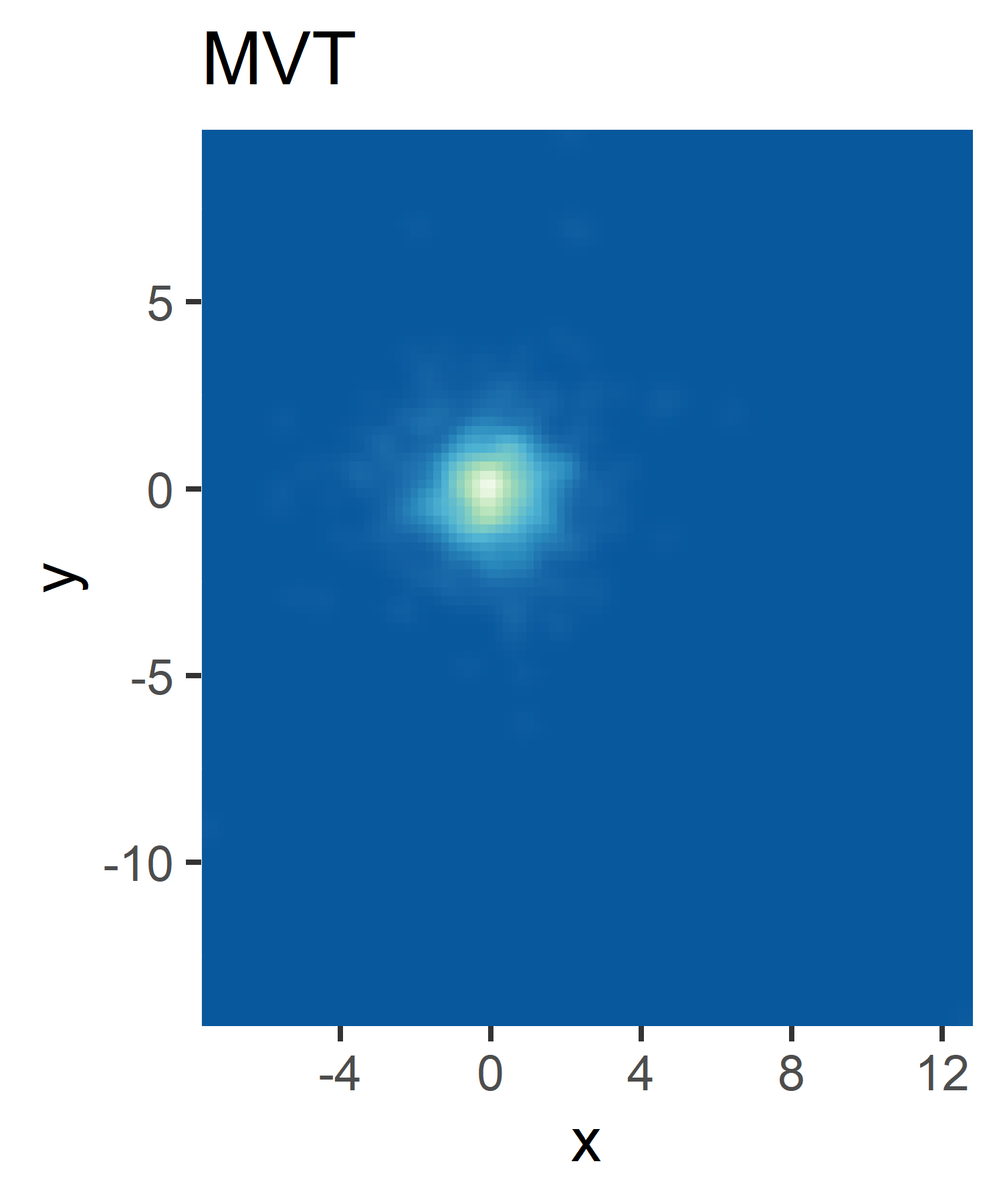}
	\includegraphics[width=0.2\textwidth]{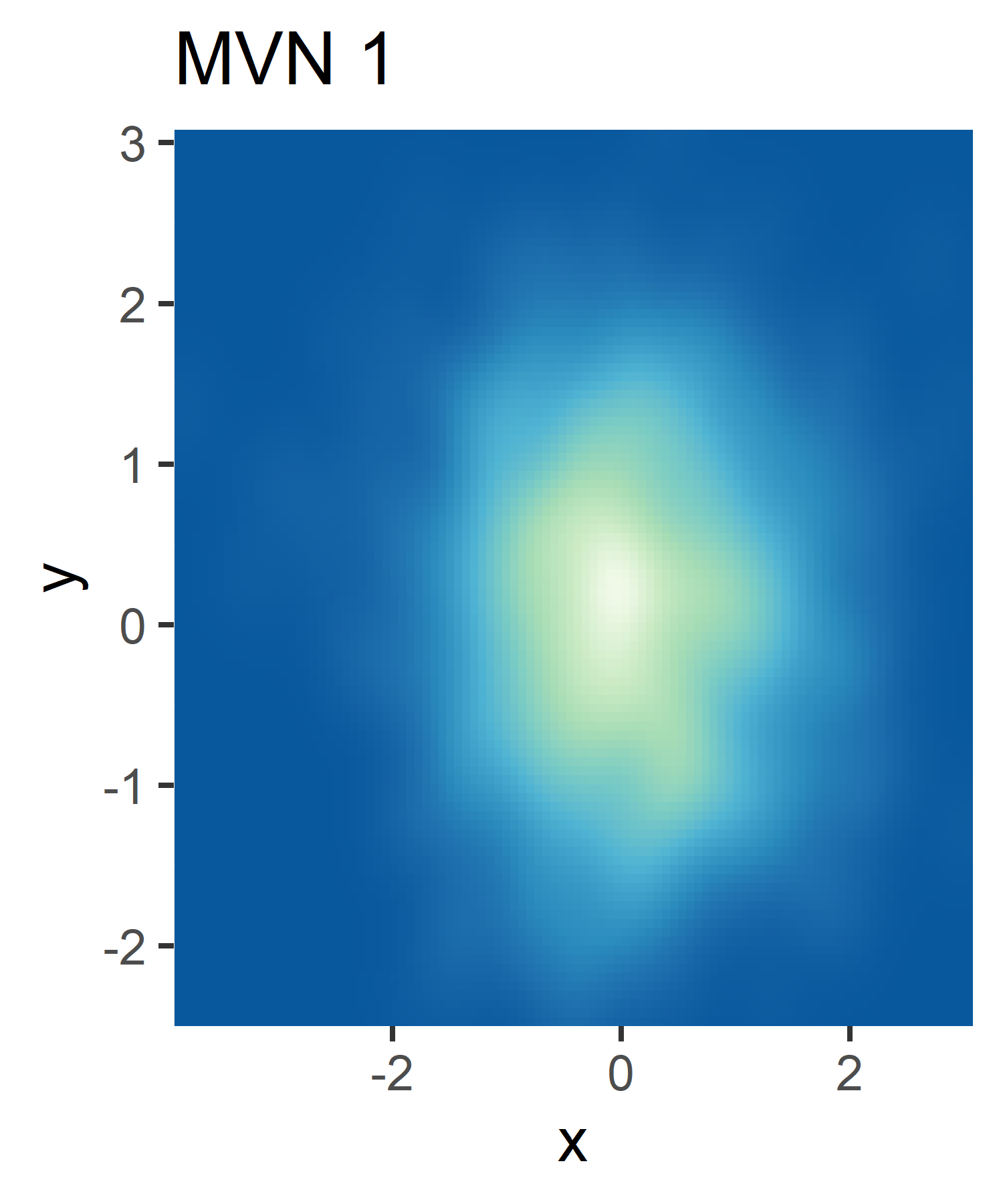}
	\includegraphics[width=0.2\textwidth]{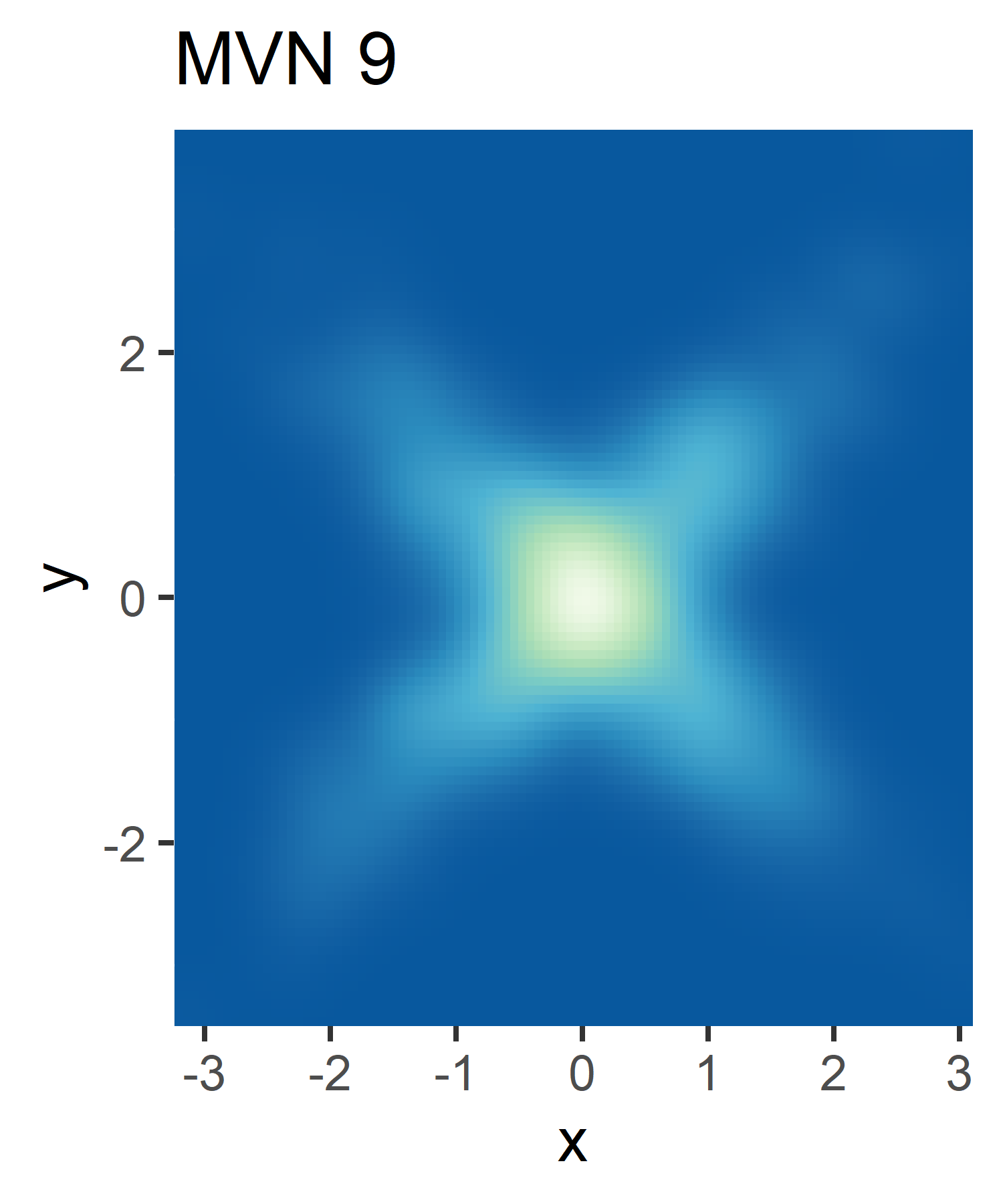}
	\includegraphics[width=0.2\textwidth]{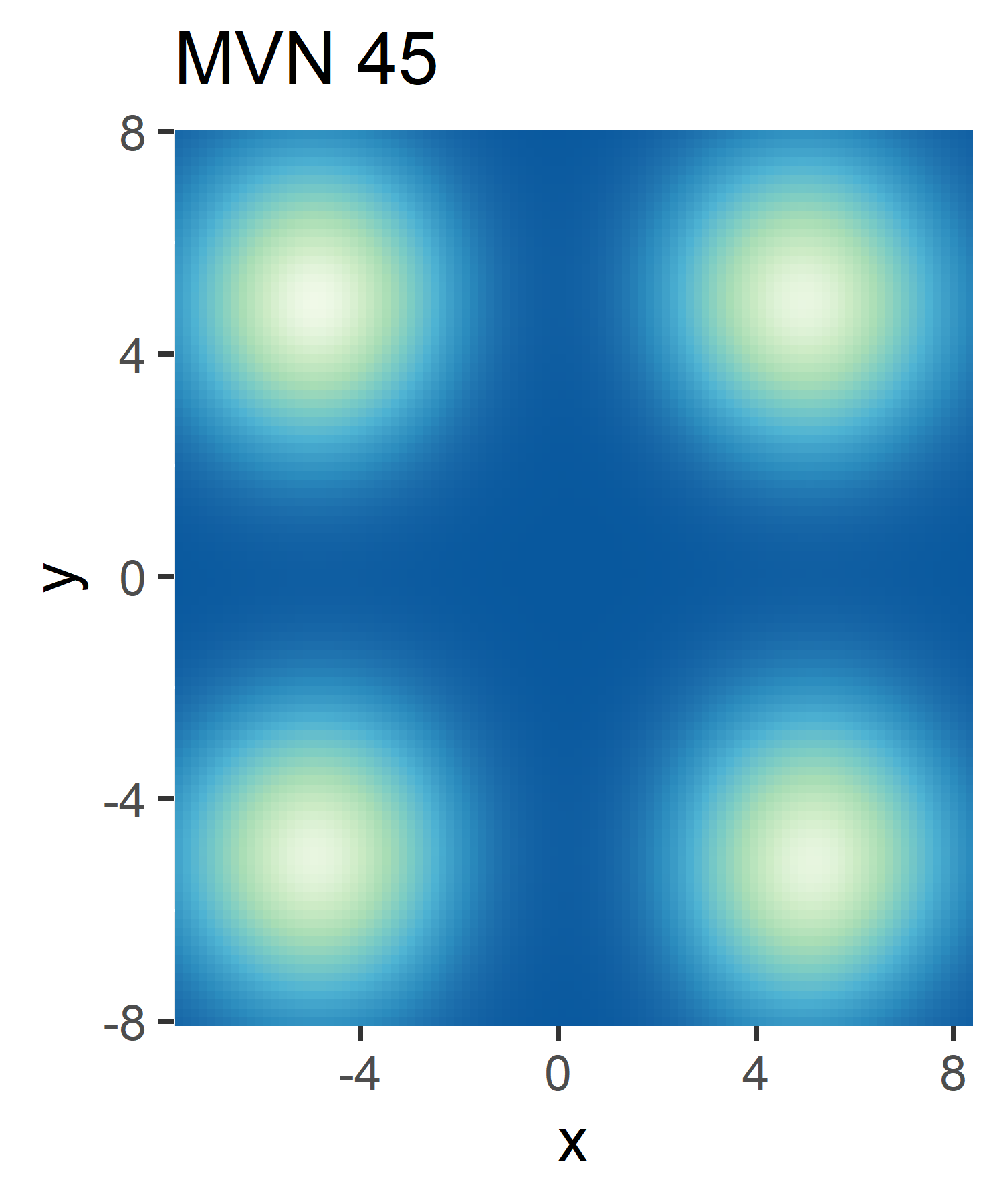}
	\caption{Density plots of the a synthetic ($n=1,000$) for each simulation distributions.}
	\label{fig:heat}
\end{figure}

\begin{figure}[h]
	\centering
	\includegraphics[width=0.95\textwidth]{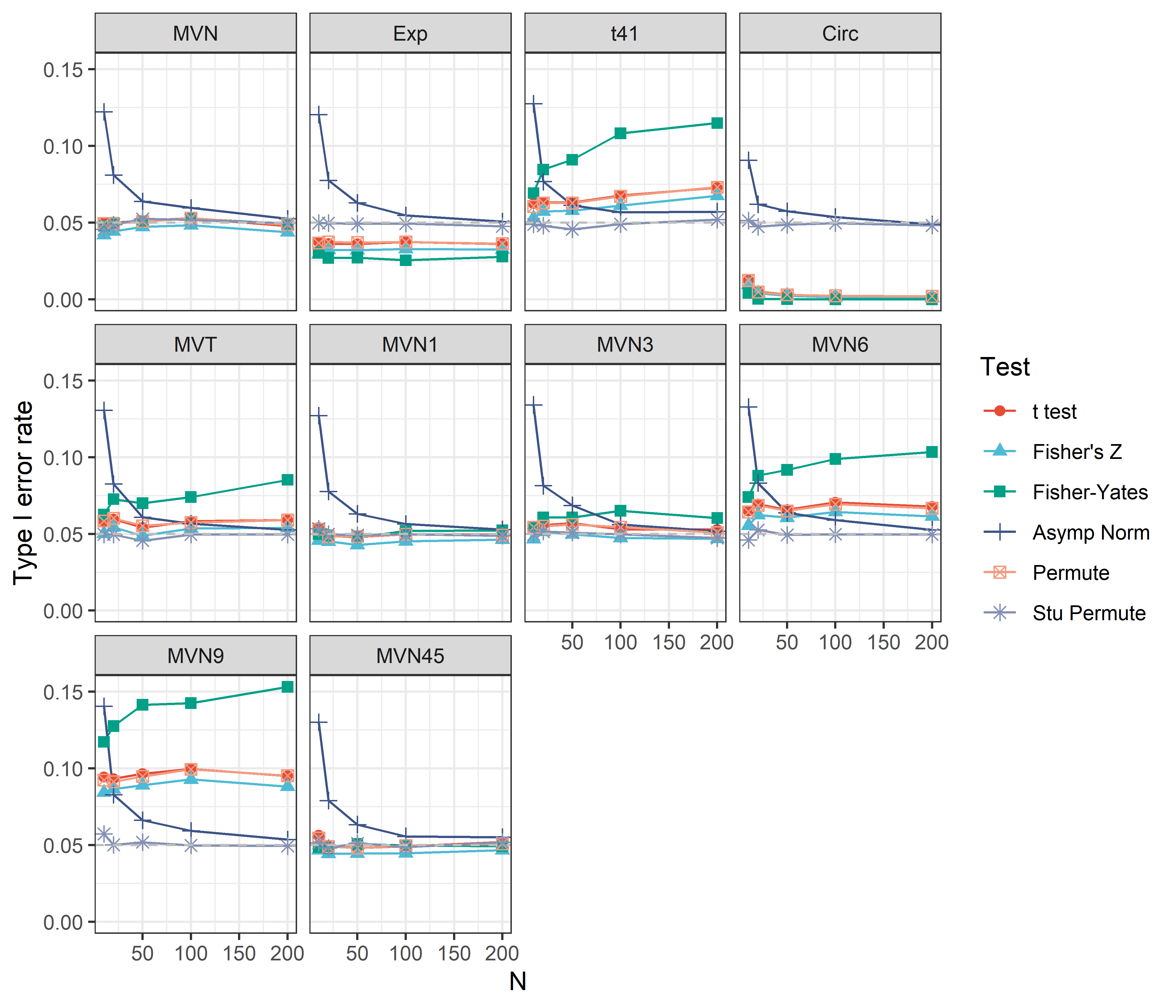}
	\caption{Type I error rate of testing $H_0: \rho_s = 0$ versus $H_0: \rho_s > 0$.}
	\label{fig:simu}
\end{figure}

The results in Table \ref{tab:simu} and Figure \ref{fig:simu} show that the large sample asymptotic normal approximation has
 inflated type I error rates for all distributions when $n \le 50$. The $t$ test, Fisher's $Z$ test, Fisher-Yates, and naive permutation tests tend to be over-conservative for the exponential and circular distributions. While for $t_{4.1}$, the type I error is consistently inflated. Note that, for these tests, such deviation cannot be corrected as sample size increases. Instead, they may converge to an arbitrary level, either lower or higher than $\alpha$. 

For MVN 1-9, we simulated a range of dependency among uncorrelated $X$ and $Y$, where MVN 1 has the weakest and MVN 9 has the strongest dependency (Figure \ref{fig:heat}). The above four tests showed the type I error rate inflation becomes increasingly severe as the dependency increases. This demonstrates the failure in controlling type I error results from the data being dependent, which can occur when the underlying distribution is non-normal.

The MVN 45 is a case where the dependency of original data is remedied by using the ranks. In this case, the ranks will distribute as if it comes from a bivariate normal distribution, regardless of the distance between the centers of individual Gaussian sub-populations. Therefore, all four tests show well control of the type I error rate.

On the other hand, the studentized permutation test robustly control type I error for all distributions examined, even when the $n$ is as small as 10. This demonstrates a clear advantage of the proposed test over all other commonly used tests for Spearman's correlation coefficient.

\begin{table}[!htbp] \centering 
	\caption{Type I error rate of testing $H_0: \rho_s = 0$ versus $H_0: \rho_s > 0$.} 
	\label{tab:simu} 
	\begin{tabular}{@{\extracolsep{5pt}} rrrrrrrr} 
		\\[-1.8ex]\hline 
		\hline \\[-1.8ex] 
		Distribution & N & $t$ test & Fisher's $Z$ & Fisher-Yates & Asymp Norm & Permute & Stu Permute \\ 
		\hline \\[-1.8ex] 
		MVN & $10$ & $0.0498$ & $0.0420$ & $0.0479$ & $0.1223$ & $0.0495$ & $0.0457$ \\ 
		 & $20$ & $0.0501$ & $0.0444$ & $0.0499$ & $0.0811$ & $0.0494$ & $0.0487$ \\ 
		 & $50$ & $0.0512$ & $0.0473$ & $0.0509$ & $0.0639$ & $0.0508$ & $0.0525$ \\ 
		 & $100$ & $0.0529$ & $0.0484$ & $0.0527$ & $0.0596$ & $0.0530$ & $0.0518$ \\ 
		 & $200$ & $0.0479$ & $0.0438$ & $0.0493$ & $0.0525$ & $0.0487$ & $0.0496$ \\
		 \hline \\[-1.8ex]  
		Exponential & $10$ & $0.0375$ & $0.0295$ & $0.0303$ & $0.1205$ & $0.0368$ & $0.0498$ \\ 
		 & $20$ & $0.0363$ & $0.0323$ & $0.0270$ & $0.0776$ & $0.0375$ & $0.0497$ \\ 
		 & $50$ & $0.0362$ & $0.0321$ & $0.0272$ & $0.0629$ & $0.0371$ & $0.0493$ \\ 
		 & $100$ & $0.0374$ & $0.0329$ & $0.0255$ & $0.0547$ & $0.0374$ & $0.0495$ \\ 
		 & $200$ & $0.0362$ & $0.0325$ & $0.0278$ & $0.0507$ & $0.0364$ & $0.0476$ \\ 
		 \hline \\[-1.8ex] 
		$t_{4.1}$ & $10$ & $0.0625$ & $0.0531$ & $0.0693$ & $0.1275$ & $0.0606$ & $0.0489$ \\ 
		 & $20$ & $0.0631$ & $0.0574$ & $0.0846$ & $0.0768$ & $0.0629$ & $0.0480$ \\ 
		 & $50$ & $0.0631$ & $0.0579$ & $0.0910$ & $0.0614$ & $0.0628$ & $0.0457$ \\ 
		 & $100$ & $0.0677$ & $0.0611$ & $0.1082$ & $0.0567$ & $0.0672$ & $0.0491$ \\ 
		 & $200$ & $0.0728$ & $0.0675$ & $0.1149$ & $0.0572$ & $0.0731$ & $0.0521$ \\ 
		 \hline \\[-1.8ex] 
		Circular & $10$ & $0.0124$ & $0.0095$ & $0.0041$ & $0.0907$ & $0.0124$ & $0.0514$ \\ 
		 & $20$ & $0.0052$ & $0.0040$ & $0.0004$ & $0.0620$ & $0.0048$ & $0.0474$ \\ 
		 & $50$ & $0.0031$ & $0.0024$ & $0.0001$ & $0.0575$ & $0.0030$ & $0.0487$ \\ 
		 & $100$ & $0.0021$ & $0.0015$ & $<0.0001$ & $0.0537$ & $0.0024$ & $0.0497$ \\ 
		 & $200$ & $0.0016$ & $0.0009$ & $<0.0001$ & $0.0489$ & $0.0020$ & $0.0482$ \\
		 \hline \\[-1.8ex]  
		MVT & $10$ & $0.0582$ & $0.0502$ & $0.0626$ & $0.1305$ & $0.0584$ & $0.0497$ \\ 
		 & $20$ & $0.0596$ & $0.0543$ & $0.0726$ & $0.0825$ & $0.0602$ & $0.0496$ \\ 
		 & $50$ & $0.0542$ & $0.0489$ & $0.0700$ & $0.0608$ & $0.0553$ & $0.0456$ \\ 
		 & $100$ & $0.0584$ & $0.0535$ & $0.0740$ & $0.0566$ & $0.0574$ & $0.0496$ \\ 
		 & $200$ & $0.0590$ & $0.0539$ & $0.0852$ & $0.0527$ & $0.0590$ & $0.0496$ \\ 
		 \hline \\[-1.8ex] 
		MVN 1 & $10$ & $0.0541$ & $0.0458$ & $0.0499$ & $0.1271$ & $0.0531$ & $0.0534$ \\ 
		 & $20$ & $0.0484$ & $0.0451$ & $0.0490$ & $0.0777$ & $0.0482$ & $0.0498$ \\ 
		 & $50$ & $0.0481$ & $0.0428$ & $0.0478$ & $0.0631$ & $0.0484$ & $0.0492$ \\ 
		 & $100$ & $0.0497$ & $0.0452$ & $0.0520$ & $0.0564$ & $0.0498$ & $0.0497$ \\ 
		 & $200$ & $0.0495$ & $0.0462$ & $0.0523$ & $0.0529$ & $0.0493$ & $0.0488$ \\ 
		 \hline \\[-1.8ex] 
		MVN 3 & $10$ & $0.0547$ & $0.0467$ & $0.0539$ & $0.1341$ & $0.0544$ & $0.0497$ \\ 
		 & $20$ & $0.0557$ & $0.0514$ & $0.0607$ & $0.0814$ & $0.0551$ & $0.0516$ \\ 
		 & $50$ & $0.0571$ & $0.0498$ & $0.0607$ & $0.0686$ & $0.0562$ & $0.0508$ \\ 
		 & $100$ & $0.0530$ & $0.0473$ & $0.0651$ & $0.0561$ & $0.0544$ & $0.0497$ \\ 
		 & $200$ & $0.0529$ & $0.0468$ & $0.0603$ & $0.0517$ & $0.0508$ & $0.0473$ \\ 
		 \hline \\[-1.8ex] 
		MVN 6 & $10$ & $0.0648$ & $0.0555$ & $0.0740$ & $0.1328$ & $0.0643$ & $0.0461$ \\ 
		 & $20$ & $0.0694$ & $0.0624$ & $0.0880$ & $0.0832$ & $0.0686$ & $0.0526$ \\ 
		 & $50$ & $0.0658$ & $0.0606$ & $0.0917$ & $0.0638$ & $0.0650$ & $0.0494$ \\ 
		 & $100$ & $0.0706$ & $0.0644$ & $0.0988$ & $0.0590$ & $0.0694$ & $0.0498$ \\ 
		 & $200$ & $0.0678$ & $0.0615$ & $0.1034$ & $0.0527$ & $0.0667$ & $0.0496$ \\ 
		 \hline \\[-1.8ex] 
		MVN 9 & $10$ & $0.0944$ & $0.0843$ & $0.1173$ & $0.1405$ & $0.0924$ & $0.0573$ \\ 
		 & $20$ & $0.0932$ & $0.0865$ & $0.1276$ & $0.0829$ & $0.0913$ & $0.0502$ \\ 
		 & $50$ & $0.0965$ & $0.0891$ & $0.1414$ & $0.0662$ & $0.0948$ & $0.0519$ \\ 
		 & $100$ & $0.0997$ & $0.0928$ & $0.1425$ & $0.0593$ & $0.0994$ & $0.0498$ \\ 
		 & $200$ & $0.0950$ & $0.0882$ & $0.1531$ & $0.0536$ & $0.0953$ & $0.0494$ \\ 
		 \hline \\[-1.8ex] 
		MVN 45 & $10$ & $0.0564$ & $0.0467$ & $0.0487$ & $0.1301$ & $0.0547$ & $0.0515$ \\ 
		 & $20$ & $0.0492$ & $0.0444$ & $0.0480$ & $0.0791$ & $0.0491$ & $0.0482$ \\ 
		 & $50$ & $0.0481$ & $0.0445$ & $0.0507$ & $0.0634$ & $0.0482$ & $0.0512$ \\ 
		 & $100$ & $0.0492$ & $0.0447$ & $0.0498$ & $0.0557$ & $0.0497$ & $0.0488$ \\ 
		 & $200$ & $0.0521$ & $0.0468$ & $0.0492$ & $0.0552$ & $0.0509$ & $0.0519$ \\ 
		\hline \\[-1.8ex] 
	\end{tabular} 
\end{table}

\section{Application}
\label{sec:application}

\subsection{TCGA breast cancer data}
\begin{figure}[!htbp]
	\centering
	\includegraphics[width=0.3\textwidth]{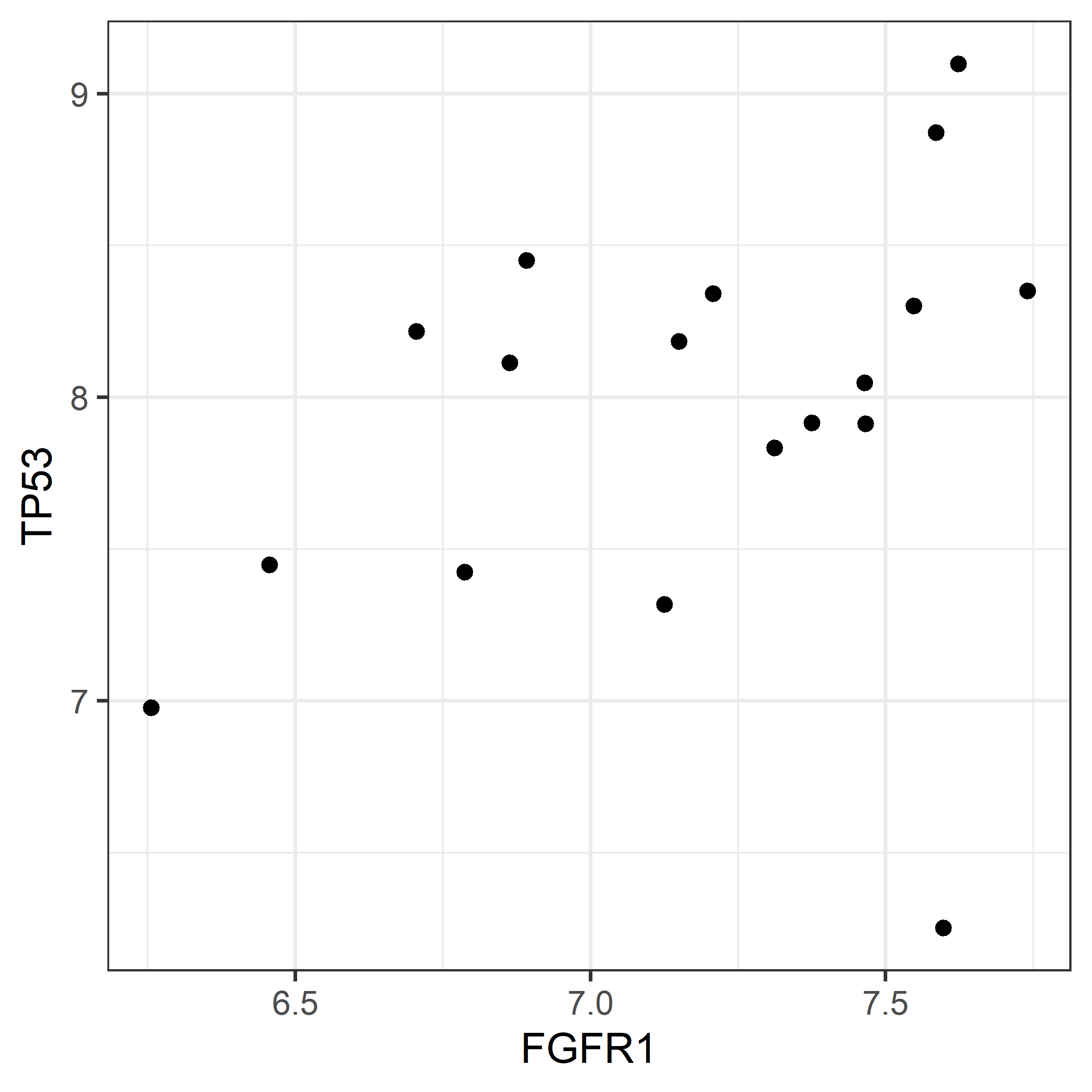}
	\includegraphics[width=0.3\textwidth]{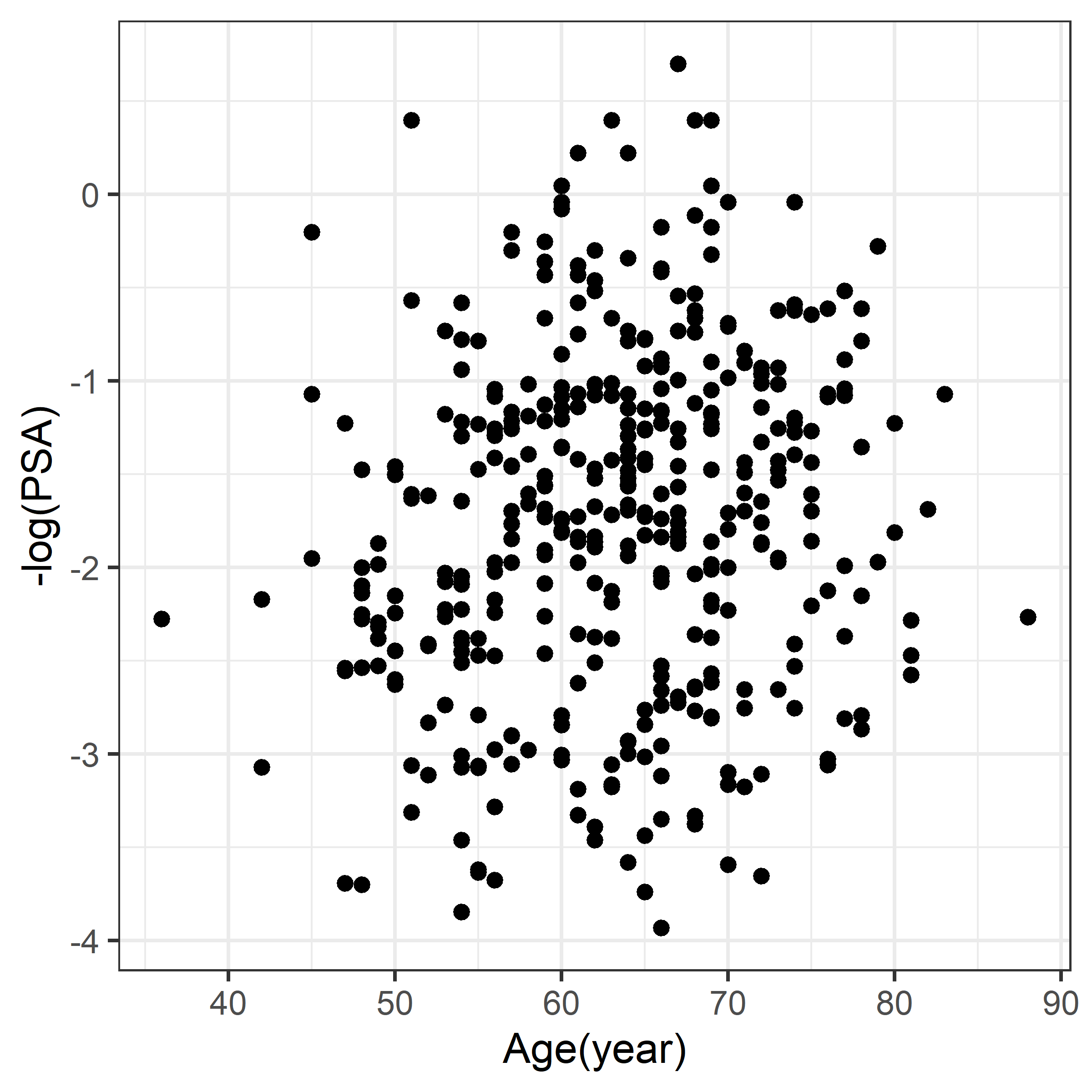}
	\caption{Scatter plot of log-transformed TP53 versus FGFR1 abundance  for TCGA data (\textit{left}), and age versus $-\log(\text{PSA})$ for the PSA data (\textit{right}).}
	\label{fig:tcga_1}
\end{figure}

As an illustration of our approach, we tested $H_0: \rho_s = 0$ versus $H_1: \rho_s > 0 $ using The Cancer Genome Atlas (TCGA) breast cancer RNA sequencing (RNA-seq) data. The gene abundance was RSEM normalized \cite{li2011rsem}. Fibroblast growth factor (FGF)2, FGF4, FGF7 and FGF20 are representative paracrine FGFs binding to heparan-sulfate proteoglycan and fibroblast growth factor receptors (FGFRs), whereas FGF19, FGF21 and FGF23 are endocrine FGFs binding to Klotho and FGFRs. FGFR1 is relatively frequently amplified and overexpressed in breast and lung cancer, and FGFR2 in gastric cancer. Moreover, FGF2 activates human dermal fibroblasts through transcriptional downregulation of the TP53 gene \cite{katoh2016fgfr}. In this application, we examine whether the transcriptomic abundance of FGFR1 is correlated with that of TP53. To investigate the performance in small sample settings, we selected 18 samples from 17 mucinous carcinoma patients. The scatter plot of log-transformed TP53 and FGFR1 abundances is shown in Figure \ref{fig:tcga_1} (\textit{left}). The marginal normality of data was examined by Shapiro-Wilk test and the bivariate normality was examined by Henze-Zikler test. The $p$ values of Shapiro-Wilk tests for log-transformed TP53 and FGFR1 abundances are 0.6077 and 0.0644, respectively. The $p$ value of Henze-Zikler test is 0.3478. Although there is no statistical significant, the marginal of FGFR1 abundance likely deviates from normal distribution.

The estimated Spearman's correlation is $\hat{\rho}_s=0.4056$. Table \ref{tab:tcga_1} shows the results of hypothesis testing. Only the result of studentized permutation test is non-significant at $\alpha=0.05$ and suggests there is no evidence of positive correlation between TP53 and FGFR1. In fact the biology does not support a positive correlation either, since FGFR1 mediates negative regulation of TP53 by FGF2 at transcriptional level \cite{katoh2016fgfr}. Indeed, if we include all samples ($n\approx 1200$) from TCGA breast cancer cohort, then all tests will fail to reject $H_1$ with $p$-values over 0.5 except for Fisher-Yates test. Together with the results from the simulations, the result by studentized permutation test is clearly more reliable. 

\begin{table}[!htbp] \centering 
	\caption{Results of testing $H_0: \rho_s = 0$ versus $H_1: \rho_s > 0$ for TCGA breast cancer data and PSA data.} 
	\label{tab:tcga_1} 
	\begin{tabular}{@{\extracolsep{5pt}} lrr} 
		\hline \\[-1.8ex] 
		Tests & $p$ value (TCGA) & $p$ value (PSA) \\ 
		\hline \\[-1.8ex] 
		$t$ test & $0.047$ & $<0.001$\\ 
		Fisher's $Z$ & $<0.001$ & $<0.001$\\ 
		Fisher-Yates & $<0.001$ & $<0.001$\\
		Asymp Norm & $0.039$ & $<0.001$ \\
		Permute & $0.033$ & $<0.001$ \\ 
		Stu Permute & $0.081$ & $<0.001$ \\ 
		\hline \\[-1.8ex] 
	\end{tabular} 
\end{table} 

\subsection{PSA data}
The testing methods were also applied to a data set of age and baseline prostate-specific antigen (PSA) levels \cite{sweeney2015chemohormonal}. The data consists of age and PSA levels of 480 subjects, of which 473 have complete paired observations. The sample Spearman's correlation coefficient between age and PSA is $\hat{\rho}_s=-0.1622$. Since the alternative hypothesis of proposed test is $H_1: \rho_s>0 $, we applied a negative log transformation on PSA levels. Similar as the TCGA example, the marginal normality of data was examined by Shapiro-Wilk test, and the bivariate normality was examined by Henze-Zikler test. The $p$ values of Shapiro-Wilk tests for log-transformed age and PSA levels are 0.0208 and 0.0301, respectively. The $p$ value of Henze-Zikler test is $<0.0001$. The results indicates the distribution is not bivariate normal. Figure \ref{fig:tcga_1} (\textit{right}) shows the scatter plot of age versus $-\log(\text{PSA})$. Table \ref{tab:tcga_1} shows that all tests rejects the $H_1$ and conclude there is a non-zero correlation between age and PSA. The is an example where all tests have consistent results. Although the normality tests are significant, such deviation may have been remedied by using the ranks in this specific example.

\section{Discussion}
\label{sec:discussion}

Conventional tests of the Spearman's correlation rely on normality assumption, including $t$-test, Fisher's $Z$ transformation, and naive permutation test, which fails to control Type I error rates when the assumption is violated. This was illustrated in our simulations studies (Section \ref{sec:simulations}). Such defect cannot be remedied by transforming the marginal distributions such as by Fisher-Yates coefficient. Notably, the deviation from bivariate normality can result in a convergence of type I error rate to an arbitrary level when $n\to \infty$. This indicates that, under scenarios when two random variables are uncorrelated but dependent, the type I error will not be controlled at desired level no matter how large the sample size is.  On the other hand, the Serfling's test based on delta method guarantees that the type I error rate converges to $\alpha$ as long as the fourth order moment is finite. However, it typically suffers an inflated type I error when sample size is under 50. 

In this work, we present a robust Spearman's correlation permutation test based on studentized statistic for testing $H_0: \rho_s = 0$ versus $H_1: \rho_s > 0$. The proposed approach is inspired by the work by DiCiccio and Romano \cite{diciccio2017robust}, which was developed for Pearson's correlation. Through extensive simulation studies and real world application, we show the proposed test controls type I error even when sample size is as small as 10 and normality assumption is violated. Therefore, the test is valid in general cases. In addition, the studentized statistic can also be used for bootstrapping tests, so as to test for more general point null hypotheses \cite{hutson2019robust}. In conclusion, the proposed studentized permutation test should be used as a routine for testing non-zero Spearman's correlation coefficient.

\section*{Acknowledgments}
This work was supported by Roswell Park Cancer Institute and National Cancer Institute (NCI) grant P30CA016056,  NRG Oncology Statistical and Data Management Center grant U10CA180822 and IOTN Moonshot grant U24CA232979-01. The results shown here are in part based upon data generated by the TCGA Research Network: https://www.cancer.gov/tcga. The PSA data example is based on research using information obtained from www.projectdatasphere.org, which is maintained
by {\em Project Data Sphere, LLC}. Neither {\em Project Data Sphere, LLC} 
nor the owner(s) of any information from the website
have contributed to, approved or are in any way responsible for the contents of this publication.

\bibliographystyle{unsrt}
\bibliography{references}  

\end{document}